\documentclass[aip,showpacs,superscriptaddress,onecolumn,preprint]{revtex4}%
\usepackage{amsfonts}
\usepackage{amsmath}
\usepackage{amssymb}
\usepackage{graphicx}%
\begin{document}
\title{Thermophysical properties of liquid carbon dioxide under shock compressions: Quantum molecular dynamic simulations}
\author{Cong Wang}
\affiliation{LCP, Institute of Applied Physics and Computational
Mathematics, P.O. Box 8009, Beijing 100088, People's Republic of
China}
\author{Ping Zhang}
\thanks{Corresponding author; zhang\underline{
}ping@iapcm.ac.cn} \affiliation{LCP, Institute of Applied Physics
and Computational Mathematics, P.O. Box 8009, Beijing 100088,
People's Republic of China}
\affiliation{Center for Applied Physics
and Technology, Peking University, Beijing 100871, People's Republic
of China}

\begin{abstract}
Quantum molecular dynamic simulations are introduced to study the
dynamical, electrical, and optical properties of carbon dioxide
under dynamic compressions. The principal Hugoniot derived from the
calculated equation of states is demonstrated to be well accordant
with experimental results. Molecular dissociation and recombination
are investigated through pair correlation functions, and
decomposition of carbon dioxide is found to be between 40 and 50 GPa
along the Hugoniot, where nonmetal-metal transition is observed. In
addition, the optical properties of shock compressed carbon dioxide
are also theoretically predicted along the Hugoniot.

\end{abstract}

 \maketitle

\section{INTRODUCTION}
\label{sec-introduction}

Knowledge of pressure-induced transformations for materials under
extreme conditions, which requires accurate understandings of the
thermophysical properties into new and complex region, has gained
much interest in a large number of scientific and technological
domains \cite{PBX:Ernstorfer:2009}. Carbon dioxide (CO$_{2}$) is one
kind of the most important molecules in planetary science and
chemical (or explosive) process. For instance, the magnetic fields
and evolution of planets are closely related to the equation of
states (EOS) and electrical properties of planetary interiors
\cite{PBX:Elert:2003,PBX:Stevenson:1982}, which are composed of
species containing C, H, N, O, etc.; Chemical reactions of
carbonaceous molecules (such as CO$_{2}$, CO), nitrogen (N$_{2}$),
oxygen (O$_{2}$) and their mixture at high densities and
temperatures are focused due to their presence in reacted chemical
explosives
\cite{PBX:Armstrong:2009,PBX:Baber:1974,PBX:Davies:1965,PBX:Zhang:2005}.
From these aspects, clearly, an accurate description of
thermodynamical properties of CO$_{2}$ at high pressures and
temperatures is fundamental and proves itself to be indispensabe for
developing realistic models of planetary interiors and reacted
explosives.

Solid phase diagram of carbon dioxide up to 80 GPa and
molecular-nonmolecular transition have already been probed through
diamond-anvil cells and confocal micro-Raman spectroscopy
\cite{PBX:Giordano:2006,PBX:Tschauner:2001,PBX:Montoya:2008,PBX:Sengupta:2009}.
Solid specimens were reported to be stable until the formation of
\emph{a}-carbonia (an amorphous extended phase), which can only be
revealed around 50 GPa at elevated temperature or above 65 GPa at
ambient temperature \cite{PBX:Sengupta:2009}. Here, we concentrate
on liquid specimens, which were first single shocked to 29 GPa and
double-shocked to 56 GPa by Schott \cite{PBX:Schott:1991}, then
single shocked to 71 GPa by Nellis \emph{et al.}
\cite{PBX:Nellis:1991} using a two-stage light-gas gun.
Theoretically, intermolecular potential method was applied to study
the EOS of CO$_{2}$ under dynamic compressions, where the onset of
decomposition was reported to be around 30 GPa (4500 K, 17
cm$^{3}$/mol) along the Hugoniot \cite{PBX:Nellis:1991}. Despite
that the predicted EOS were in accord with experimental data, the
electronic structure, which has been proved to be important in
determining the dynamical, electrical, and optical properties of
molecular fluids under extreme conditions
\cite{PBX:Bastea:2001,PBX:Wang:2010a,PBX:Wang:2010b}, is still
lacking due to the intrinsic approximations of this method.

On the other hand, quantum molecular dynamic (QMD) simulations,
where quantum-mechanical treatments are executed by combining
classical molecular dynamics for the ions and density functional
theory (DFT) for electrons, have already been proved to be
successful in describing the thermodynamical properties of mono
atomic molecules (He) \cite{PBX:Militzer:2006}, diatomic molecules
(H$_{2}$, N$_{2}$, O$_{2}$)
\cite{PBX:Recoules:2009,PBX:Mazevet:2003,PBX:Recoules:2003}, and
hydrocarbons (C$_{6}$H$_{6}$) \cite{PBX:Wang:2010b} under extreme
conditions. A fully quantum-mechanical description of CO$_{2}$ under
shock compressions is highly recommended to be presented and
understood because of the following aspects: (i) CO$_{2}$ undergos
dramatic transformations under shock compressions, such as
dissociation and recombination of molecules, during which
carbonaceous molecules combined with oxygen (O$_{2}$) might be
produced, and electron spin polarization should be seriously
considered at this stage due to the fact that spin-triplet state
governs the ground state of oxygen \cite{PBX:Sugimori:2007}; (ii)
Electrical conductivity of molecular fluid, where electronic
structure is of dominance, can be greatly influenced by
thermo-activated electronic states under extreme conditions; (iii)
Temperature, which is an important parameter in experimental
determination of EOS, is usually difficult to be measured because of
the uncertainty in determining the optical intensity for ultraviolet
part of the spectrum, and QMD simulations are powerful tools to
provide predictions.

In the present work, DFT-based QMD simulations have been used to
investigate thermophysical properties of CO$_{2}$ under extreme
conditions. The EOS and pair correlation functions (PCF) are
determined through QMD simulations at equilibrium. Dynamic
conductivity $\sigma(\omega)$ is calculated by Kubo-Greenwood
formula, from which the dc conductivity ($\sigma_{dc}$) is
determined. Then, the dielectric function $\epsilon(\omega)$ and
reflectivity are extracted. The rest of this paper is organized as
follows. The simulation details are briefly described in Sec.
\ref{sec-method}; The PCF, which is used to study the dissociation
of CO$_{2}$, and the Hugoniot curve are given in Sec. \ref{sec-eos};
In Sec. \ref{sec-dynamic}, nonmetal-metal transition and optical
properties are discussed. Finally, we close our paper with a summary
of our main results in Sec. \ref{sec-conclusion}.

\section{COMPUTATIONAL METHOD}
\label{sec-method}

The Vienna Ab-initio Simulation Package (VASP)
\cite{PBX:Kresse:1993,PBX:Kresse:1996}, which was developed at the
Technical University of Vienna, has been employed to perform
simulations for carbon dioxide. The elements of our calculations
consist of a series of volume-fixed supercells including $N$ atoms,
which are repeated periodically throughout the space. By involving
Born-Oppenheimer approximation, electrons are quantum mechanically
treated through plane-wave, finite-temperature (FT) DFT
\cite{PBX:Lenosky:2000}, where the electronic states are populated
according to Fermi-Dirac distributions at temperature $T_{e}$. The
exchange-correlation functional is determined by generalized
gradient approximation (GGA) with the parametrization of Perdew-Wang
91 \cite{PBX:Perdew:1991}. The ion-electron interactions are
represented by a projector augmented wave (PAW) pseudopotential
\cite{PBX:Blochl:1994}. Isokinetic ensemble (NVT) is adopted in
present simulations, where the ionic temperature $T_{i}$ is
controlled by No\'{s}e thermostat \cite{{PBX:Nose:1984}}, and the
system is kept in local thermodynamical equilibrium by setting the
electron ($T_{e}$) and ion ($T_{i}$) temperatures to be equal.
Electron spin polarization has also been taken into account, due to
the possible existence of oxygen molecules in shocked CO$_{2}$.

The plane-wave cutoff energy is selected to be 600 eV so that the
pressure is converged within 5\% accuracy. $\Gamma$ point and
4$\times$4$\times$4 Monkhorst-Pack scheme \textbf{k} points are used
to sample the Brillouin zone in molecular dynamics simulation and
electronic structure calculation, respectively, because EOS
(conductivity) can only be modified within 5\% (15\%) for the
selection of higher number of \textbf{k} points. 81 atoms (27
CO$_{2}$ molecules) are included in the cubic supercell. The
densities selected in our simulations range from 13 to 37.55
cm$^{3}$/mol and temperatures between 218 and 10000 K, which
highlight the regime of the Hugoniot. All the dynamic simulations
are lasted for 3 $\sim$ 6 ps, and the time steps for the
integrations of atomic motion are 0.5 $\sim$ 1 fs according to
different densities (temperatures). Then, the subsequent 1 ps
simulations are used to calculate EOS as running averages.

\section{RESULTS AND DISCUSSION}
\label{sec-analysis}

\subsection{EQUATION OF STATE}
\label{sec-eos}

\begin{figure}[ptb]
\centering
\includegraphics{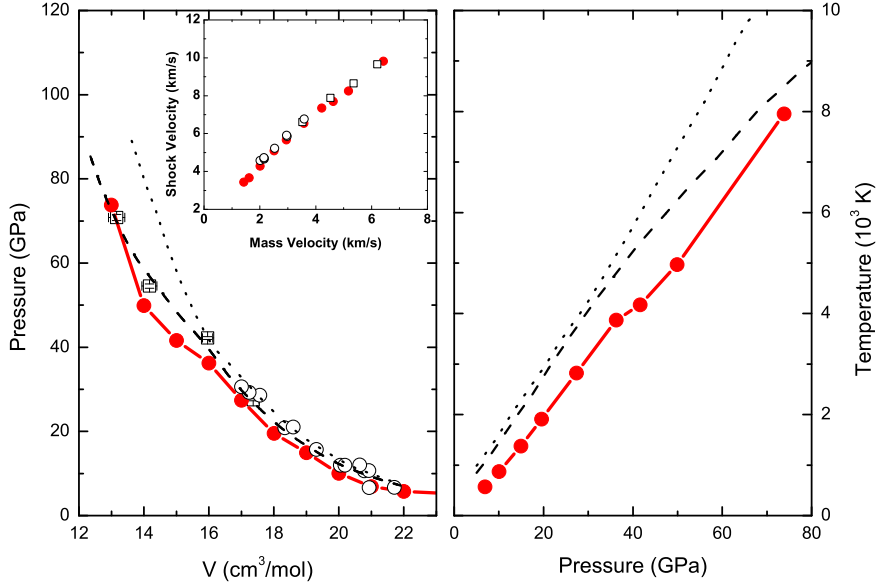}\caption{(Color online) The $P$-$V$ (left panel) and $T$-$P$ (right panel)
Hugoniot curves of shocked carbon dioxide. Inset is the ($u_{s}$,
$u_{p}$) diagram. Here, filled red circles denote the present QMD
results, while the dashed and dotted lines denote the intermolecular
potential results \cite{PBX:Nellis:1991} with and without accounting
for molecular dissociation, respectively. The experimental data
reported in Ref. \cite{PBX:Schott:1991} (open circles) and Ref.
\cite{PBX:Nellis:1991} (open squares) are also
shown for comparison. }%
\label{fig_hugoniot}%
\end{figure}

Accurate understandings of the electrical and optical properties
of CO$_{2}$ depend on a precise description of dynamical
properties, such as EOS. The EOS have been examined theoretically
through the Rankine-Hugoniot equations, which follow from
conservation of mass, momentum, and energy across the front of
shock waves. The equations describe the locus of points in ($E$,
$P$, $V$)-space satisfying the relation as follows:
\begin{equation} \label{equ_hugoniot1}
E_{1}-E_{0}=\frac{1}{2}(P_{1}+P_{0})(V-V_{0}),
\end{equation}
\begin{equation} \label{equ_hugoniot2}
P_{1}-P_{0}=\rho_{0}u_{s}u_{p},
\end{equation}
\begin{equation} \label{equ_hugoniot3}
V_{1}=V_{0}(1-u_{p}/u_{s}),
\end{equation}
where $E$, $P$, $V$ denote internal energy, pressure, volume, and
subscripts 0 and 1 present the initial and shocked state,
respectively. In Eqs. (\ref{equ_hugoniot2}) and
(\ref{equ_hugoniot3}), $u_{s}$ is the velocity of the shock wave and
$u_{p}$ corresponds to the mass velocity of the material behind the
shock front. In the present work, the initial density of CO$_{2}$ is
$\rho_{0}$=1.17 g/cm$^{3}$ ($V_{0}$=37.55 cm$^{3}$/mol) and the
liquid specimen is controlled at a temperature of 218 K, where the
internal energy is $E_{0}$=$-$22.95 eV/molecule. The initial
pressure $P_{0}$ can be treated approximately as zero compared to
the high pressure of shocked states along the Hugoniot. The Hugoniot
points are obtained as follows: (i) smooth functions are used to fit
the internal energy and pressure in terms of temperature at sampled
density; (ii) then, Hugoniot points are derived from Eq.
(\ref{equ_hugoniot1}). The results are summarized in Table
\ref{H_data}.

\begin{table}[!htbp]
\centering \caption{Hugoniot pressure and temperature points derived
from QMD simulations at a series of molar volumes. }
\begin{tabular}{ccc}
\hline\hline $V$ (cm$^{3}$/mol)& $P$ (GPa) & $T$ (K)\\
\hline
22   & 5.72    & 459        \\
21   & 6.91    & 568        \\
20   & 10.05   & 873        \\
19   & 14.95   & 1374       \\
18   & 19.53   & 1909       \\
17   & 27.39   & 2821       \\
16   & 36.24   & 3873       \\
15   & 41.62   & 4175       \\
14   & 49.89   & 4968       \\
13   & 73.82   & 7955       \\
\hline\hline
\end{tabular}
\label{H_data}
\end{table}

Comparisons between our simulated Hugoniot curve and results from
experimental measurement and intermolecular potential method are
shown in Fig. \ref{fig_hugoniot} (left panel), where our results
show good agreement with experiment. The QMD simulated pressure and
temperature along the Hugoniot curve for carbon dioxide shows a
systematic behavior, except for the region with a small break in
slope around 40 to 50 GPa, where structural transitions characterize
the EOS of molecular fluid. Although the $P$-$V$ curve by
intermolecular potential method is consistent with experiments, the
predicted temperature remains too high compared with our
calculations [see Fig. \ref{fig_hugoniot} (right panel)].

\begin{figure}[ptb]
\centering
\includegraphics{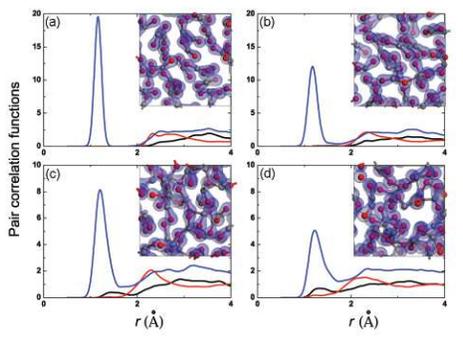}
\caption{(Color online) Calculated pair correlation function for
C-O (blue line), C-C (black line), O-O (red line) at four
densities of CO$_{2}$ along the Hugoniot. The atomic structure,
where carbon and oxygen atoms are denoted by gray and red balls
respectively, and the relative iso-surface of
charge density (blue regimes) are also provided in the insets.
(a) $V$ = 19 cm$^{3}$/mol, T = 1374 K; (b) $V$ = 16 cm$^{3}$/mol, T = 3873 K;
(c) $V$ = 15 cm$^{3}$/mol, T = 4175 K; (d) $V$ = 13 cm$^{3}$/mol, T = 7955 K.}%
\label{fig_rdf}%
\end{figure}

Dynamical properties and structural transitions of CO$_{2}$ under
extreme conditions have been examined through PCF, which represents
the possibility of finding a particle at a distance $r$ from a
reference atom. Figure \ref{fig_rdf} shows PCF and charge density
distribution (at equilibrium) of four different molar volume points,
and the structural transitions of CO$_{2}$ are clearly reflected by
PCF. At low pressures [Fig. \ref{fig_rdf}(a)], ideal molecular fluid
is suggested by PCF distribution of C-O, which exhibits a main peak
around 1.16 \AA. With the increase of pressure along the Hugoniot,
this main peak in C-O PCF reduces in amplitude and get broadened,
which is attributed to the thermo-induced vibrations of C-O bonds
around the equilibrium length. Dissociation of CO$_{2}$, which leads
to soften behavior of the Hugoniot [see Fig. \ref{fig_hugoniot}
(left panel)], has been found at about 40 to 50 GPa. It is indicated
that carbon atoms tend to form connections between each other [Fig.
\ref{fig_rdf}(c)], then, with further increase of pressure, hybrid
C-C bonds (peak around 1.45 \AA, which lies between typical
\emph{sp}$^{3}$ and \emph{sp}$^{4}$ hybrid C-C bond length),
diatomic oxygen (peak around 1.23 \AA) and mono-atomic oxygen are
formed in the shocked system [Fig. \ref{fig_rdf}(d)]. Furthermore,
we have examined the charge density difference between spin-up and
spin-down electrons, which is as small as negligible, and the
results suggest that electron spin polarization of the shock
produced mono-atomic and diatomic oxygen are suppressed at this
stage.

\subsection{DYNAMIC CONDUCTIVITY}
\label{sec-dynamic}

\begin{figure}[ptb]
\centering
\includegraphics{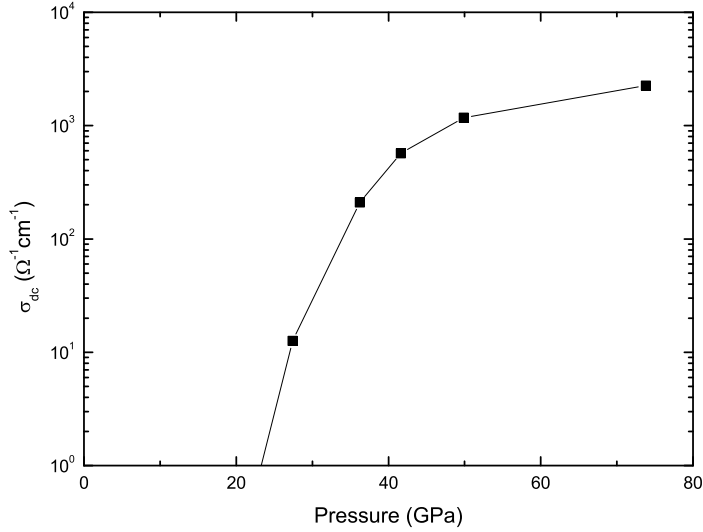}
\caption{Calculated dc conductivity of shocked carbon dioxide along the Hugoniot.}%
\label{fig_conductivity}%
\end{figure}

Great controversies have been raised since the detection of
nonmetal-metal transition of such diatomic molecules as hydrogen and
oxygen \cite{PBX:Bastea:2004,PBX:Chau:2004} in isentropic
compressions. The links between nonmetal-metal transition and
dissociation of molecules under dynamic compression are of
particular significance. Here, we examine the dynamic conductivity
of CO$_{2}$ according to Kubo-Greenwood formula:
\begin{eqnarray} \label{real-conductivity}
\sigma_{1}(\omega)&=\frac{2\pi}{3\omega\Omega}\sum\limits_{\textbf{k}}w(\textbf{k})\sum\limits_{j=1}^{N}\sum\limits_{i=1}^{N}\sum\limits_{\alpha=1}^{3}[f(\epsilon_{i},\textbf{k})-f(\epsilon_{j},\textbf{k})]
\cr
&\times|\langle\Psi_{j,\textbf{k}}|\nabla_{\alpha}|\Psi_{i,\textbf{k}}\rangle|^{2}\delta(\epsilon_{j,\textbf{k}}-\epsilon_{i,\textbf{k}}-\hbar\omega),
\end{eqnarray}
where $\Omega$ is the volume of the supercell. The $i$ and $j$
summations range over $N$ discrete bands included in the
calculation. The $\alpha$ sum is over the three spatial
directions. $f(\epsilon_{i},\textbf{k})$ describes the occupation
of the $i$th band, with the corresponding energy
$\epsilon_{i,\textbf{k}}$ and the wavefunction
$\Psi_{i,\textbf{k}}$ at $\textbf{k}$. $w(\textbf{k})$ is the
$\textbf{k}$-point weighting factor.

Along the Hugoniot, dynamic conductivity have been calculated as the
average of thirty atomic snapshots at equilibrium. The dc
conductivity ($\sigma_{dc}$), which follows the static limit
($\omega\rightarrow0$) of $\sigma_{1}(\omega)$, is then extracted
and plotted in Fig. \ref{fig_conductivity}. At pressures below 20
GPa, $\sigma_{dc}$ can be neglected, and the results suggest an
insulating molecular fluid. Thermal activation of conductivity is
raised from 20 GPa, then metallization ($\sigma_{dc} >$ 1000
$\Omega^{-1}$cm$^{-1}$) of CO$_{2}$, which is accompanied with
dissociation of molecules, is reached at 40 to 50 GPa.

\begin{figure}[ptb]
\centering
\includegraphics{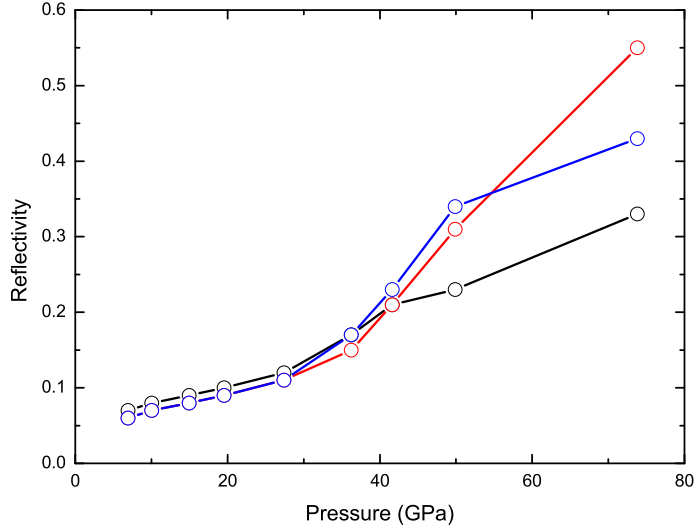}
\caption{(Color online) Optical reflectivity of shocked carbon
dioxide for wavelengths of 404 (black line), 808 (red line), and
1064 nm (blue line) along the Hugoniot.}%
\label{fig_reflectivity}%
\end{figure}

Emissivity of spectrum, which is closely related to optical
reflectivity, is important in determining temperature in experiment.
Optical reflectivity can be derived from dynamic conductivity. The
imaginary part $\sigma_{2}(\omega)$ can be obtained from the
Kramers-Kronig relation:

\begin{eqnarray} \label{ima-conductivity}
\sigma_{2}(\omega)=-\frac{2}{\pi}P\int\frac{\sigma_{1}(\nu)\omega}{\nu^{2}-\omega^{2}}d\nu,
\end{eqnarray}
where $P$ is the principal value of the integral. Then dielectric
functions can be derived from the two parts of the conductivity:
\begin{eqnarray} \label{real-dielectric}
\epsilon_{1}(\omega)=1-\frac{4\pi}{\omega}\sigma_{2}(\omega),
\end{eqnarray}
\begin{eqnarray} \label{ima-dielectric}
\epsilon_{2}(\omega)=\frac{4\pi}{\omega}\sigma_{1}(\omega).
\end{eqnarray}
The real part $n(\omega)$ and the imaginary part $k(\omega)$ of
the refraction index have the following relations with dielectric
function:
\begin{eqnarray} \label{real-refraction}
n(\omega)=\sqrt{\frac{1}{2}[|\epsilon(\omega)|+\epsilon_{1}(\omega)]},
\end{eqnarray}
\begin{eqnarray} \label{ima-refraction}
k(\omega)=\sqrt{\frac{1}{2}[|\epsilon(\omega)|-\epsilon_{1}(\omega)]}.
\end{eqnarray}
The index of refraction is useful for evaluating optical
properties such as the reflectivity $r(\omega)$ and absorption
coefficient $\alpha(\omega)$:
\begin{eqnarray} \label{reflectivity}
r(\omega)=\frac{[1-n(\omega)]^{2}+k(\omega)^{2}}{[1+n(\omega)]^{2}+k(\omega)^{2}}.
\end{eqnarray}

We have examined the optical reflectance at three wavelengths (404,
808, and 1064 nm) spanning the visible spectrum, and the results are
shown along the Hugoniot in Fig. \ref{fig_reflectivity}, which
clearly shows the pressure-induced change in reflectivity (from 0.05
to 0.35 $\sim$ 0.55). For the pressure between 20 and 40 GPa, the
increase in reflectance is retarded, then sharp increase exists
around 40 to 50 GPa due to the nonmetal-metal transition. The
results could be inspected by future experiments.

\section{CONCLUSION}
\label{sec-conclusion}

In summary, we have studied the thermodynamical properties of carbon
dioxide under extreme conditions. The EOS obtained from QMD
simulations are used to determine the Hugoniot curve, which shows
good agreement with available experimental data. Our calculations
indicate that decomposition of carbon dioxide begins at 40 to 50 GPa
along the Hugoniot, where carbon atoms intend to form connections
between each other, then diatomic and monoatomic oxygen are
produced. The nonmetal-metal transition, as well as the change in
optical reflectance, are also found in the present work.

\begin{acknowledgments}
This work was supported by NSFC under Grants No. 90921003 and No.
60776063.
\end{acknowledgments}


\end{document}